# Capillary-size Flow of Human Blood Plasma: Revealing Hidden Elasticity and Scale Dependence

U. Windberger[1], P. Baroni[2], L. Noirez[2*]

[1]Center for Biomedical Research, Decentralized Biomedical Facilities, Medical University Vienna, Borschkegasse 8a, 1090 Vienna, Austria. ORCID: 0000-0002-5800-9089

[2]Laboratoire Léon Brillouin (CEA-CNRS), Univ. Paris-Saclay, CEA-Saclay, 91191 Gif-sur-Yvette, France. *ORCID : 0000-0002-0164-1155

The dynamical mechanical analysis of blood generally uses models inspired by conventional flows, assuming scale-independent homogeneous flows and without considering fluid-surface boundary interactions. The present experimental study highlights the relevance of using an approach in line with physiological reality providing a strong interaction between the fluid and the boundary interface. New dynamic properties of human blood plasma are found: a finite shear elastic response (solid-like property) is identified in nearly static conditions, which also depends on the scale (being reinforced at small scales). The elastic behavior is confirmed by the induction of local hot and cold thermodynamic states evidencing a thermo-mechanical coupling in blood plasma so far known only in elastic materials. This finding opens new routes for medical diagnosis and device fabrication.

Keywords: Blood plasma, Interactive surface, Shear-elasticity, Flow

Introduction

The transport mechanisms of blood in the vasculature face multiple biological, physical, and biochemical parameters that make their flow laws complex, specific, and fascinating. Ninety years after the first works from Fahraeus and Lindqvist[1] showing the decrease of blood viscosity with decreasing tube diameter, characterizing physiological behavior remains a challenge. Blood behavior is not only influenced by the flow properties of blood cells, but also by their ability to couple with the surrounding plasma. Plasma is a colloidal aqueous solution having almost 10 mass% of proteins, smaller molecules, and ions[2]. It plays a key role by interacting and dragging the blood cells in the vessels. While it was long time believed to flow uncoupled (like Newtonian liquids), there is more and more evidence for non-Newtonian plasma behavior[3,4]. The flow involves the displacement of fluid layers against each other, and therefore the breaking of the fluid cohesion. This intrinsic property is rarely discussed and measured. However, recent careful tests have shown that it is possible to access the fluid cohesion at small scale probing the stress response of the fluid upon applying a mechanical shear strain. It is shown that "simple" fluids as liquid water exhibit a resistance to shear flow at small scale. The liquid elasticity vanishes by increasing the scale, giving rise to the conventional viscous behavior[5-8]. The very early experimental evidence of liquid elasticity dates back to Derjaguin (in the 1980') at few microns scales[9]. At macroscopic scale, rheology has systematized the viscoelastic study of fluids with well-established protocols based on the Maxwell model established by Frenkel in "The Kinetic Theory of Liquids" (1946)[10]. Derjaguin's works were forgotten until new measurements obstinately pointed out that the surface/liquid boundary conditions impact the mesoscale viscoelastic response[5-8]. At the same time, theoreticians have taken up Frenkel's calculation and demonstrated that fluid viscoelastic moduli do not necessarily vanish at low frequency; an elastic approach is equally valid[11-14]. The liquid shear elasticity weakens as the inverse cubic power of confinement length $L$, being rather inaccessible at macroscopic scale[12-14]. Due to the high importance of the small scales in the blood circulation, identifying scale-dependent shear elasticity and possibly coupled thermo-elastic effects[15] in human blood plasma is an utmost important challenge. To understand the underlying mechanisms and access the shear elasticity, we explored the mechanical and thermal plasma properties at different scales in relation with the microvascular dimensions.

We operated at optimized conditions to test the true fluid resistance to a dynamic mechanical shear strain, in a frequency range reproducing the heartbeat and at length scales compatible with physiological dimensions. We started from the observation that standard protocols used for viscometry or viscoelastic measurements cannot be satisfactory. Classical rheometry uses metallic surfaces that are unphysiological and the degree of "un-physiology" is not known. They do not reproduce the interactive role of endothelial lubrication that covers biological surfaces[16,17]. Classical rheometry protocol also does not consider the influence of the scale and excludes the diameters within the microvasculature. To approach true fluid properties it is necessary to adapt the dynamic mechanical measurement. We thus combine innovative conditions associating tribology and multi-scale mechanical dynamical analysis together with a simultaneous time-resolved thermo-mapping of the fluid upon strain field.

The methodology includes an optimal stress transfer through the liquid by enhancing the force of interaction of the fluid to the wall. Excellent interfacial wetting is a prerequisite to strengthen the interaction. The degree of wetting can be characterized by the method of the contact angle $\theta$ (Scheme 1). Slippage propensity is reduced when the contact angle is reduced[18]. The improved wetting method showed already that ordinary liquids (classified Newtonian by classical rheometry) exhibited a solid-like response below an elastic threshold[5,9,19,20]. We felt therefore confident to find such elasticity also in blood plasma after removal of blood cells that usually provide the elasticity[21].

Materials and Methods

We optimized the plasma – substrate interaction by means of zero-porosity α-alumina surfaces. α-alumina is chemically and biologically inert. Due to its electronic and atomic structure the α-alumina is a high-energy surface[22,23]. Blood plasma contains 90% water fulfilling the total wetting conditions (illustrated in Scheme 1). The fluid was set in the gap at ambient temperature between two α-alumina surfaces (plate-plate disks; 40 mm diameter). The gap thickness varied from 300 μm to 1100 μm. The shear strain is defined as $\gamma = \delta l/e$, where $\delta l$ is the displacement and $e$ the gap thickness. The strain values range from 1% to 2000 %, depending on the gap thickness. A sin shear strain wave $\gamma(\omega)$ was generated by the oscillatory motion of one surface while the other one is fixed and coupled to a stress sensor (using a strain imposed rheometer ARES2 (TA-Instruments)). The resulting shear stress $\sigma(\omega)$ that measures the resistance of the fluid to the deformation is conventionally expressed in terms of macroscopic viscoelastic constants to characterize the dynamic fluid properties: $\sigma(\omega) = \gamma_0.(G'(\omega).sin(\omega t) + G''(\omega).cos(\omega t))$ where ($G'$) and ($G''$) are the shear elastic and viscous moduli respectively[24]. These constants are by essence statistical quantities.

The temperature inside the fluid was simultaneously investigated using a micro bolometric detector array of 382 x 288 pixels embedded in a camera working at a frame rate of 27 Hz. The camera integrates a lens with a focal length of 7.5 mm and a numerical aperture of F/d= 0.6 detects the radiation emitted by the liquid surface in the LWIR band; i.e. wavelengths ranging between 7 and 14 μm. The thermal amplitude adjustment range is $\Delta T = 2°C$ and the thermal sensitivity +/- 0.02°C. The front lens is placed 15mm away from the liquid surface, ensuring non-contact measurements. The lens magnifies 20x. The depth of focus is defined by the pixel pitch (20μm), the lens aperture F/1 and, the distance between the object the lens pupil and the focal plane. The depth of thermal field is about 0.1mm. The images were corrected by subtraction of the equilibrium median value. The signals collected on several pixels are then added along each column (or the Y axis of the detector array). With this method, the Noise Equivalent Temperature Difference (NETD) is considerably reduced reaching typically values closed to 0.003 K. After giving informed consent (2114/2019, MedUni Vienna, Austria), blood was obtained from one healthy volunteer after given informed consent (postmenopausal female, 55a, BMI: 24, non-smoker) by puncturing the antecubital vein with a 21G butterfly needle that was connected to a vacutainer system containing EDTA for anticoagulation. The study conforms to the Declaration of Helsinki. Whole blood was centrifuged to obtain platelet-depleted plasma (2310 *g* for 15 minutes, the upper two thirds of the plasma column was withdrawn and again centrifuged at 2310 *g* for 20 minutes). The superficial layer was discarded and the central fraction was collected and immediately essayed. Blood plasma was stored at room temperature in several containers and new portions were filled in the gap for each measurement. Finally, because of the high water content (90%), blood plasma can be considered as a blackbody according to its transmission[25,26] and is therefore perfectly adapted for an infrared study.

Results

Figure 1 displays the dynamic response of a 0.300 mm thickness blood plasma sample submitted to an oscillatory shear sine input over a wide frequency range (0.04 rad/s – 100 rad/s). Low strain amplitude conditions were applied (1.5 %) to test the fluid in nearly equilibrium conditions. The elastic modulus $G'$ dominated the dynamic response indicating that plasma responds nearly instantaneously to the oscillatory strain field within this frequency. Viscous dissipation was low, since $G''$ exhibited values three times lower than $G'$. This means that plasma did not flow but oscillated as a whole in phase with the shear excitation. Both moduli increased at higher frequency, indicating a reinforcement of the strength at faster motions.

Second, the response of blood plasma was examined as a function of the strain amplitude at different thicknesses. Figure 2 displays the evolution of the shear elasticity as a function of the strain amplitude at small scale (0.300 mm) with the corresponding thermal mapping displayed at three shear strain amplitudes. Three dynamic regimes are distinguished. At relatively low strain amplitude ($\gamma < 200$ %), the shear elastic modulus $G'(\omega)$ was higher than the shear viscous modulus $G''(\omega)$. At higher strain amplitudes (300 % $< \gamma <$ 500 %), both moduli collapsed with a more pronounced fall for $G'(\omega)$, leading to a crossover, which marks the change of behavior from solid-like to viscous-like. At very high strain amplitude ($\gamma >$ 600 %), the viscous modulus dominated and the elastic component became negligible (less than 1 % of $G''(\omega)$). Viscous behavior was recovered in the high strain amplitude regime only. Prior to it, blood plasma exhibited an elastic-like regime.

Similar features could be found in the thermal behavior (Figure 2). At low strain amplitude, the temperature was relatively homogeneous. As strain amplitudes increase, thermal bands of opposite temperatures appeared along the strain direction (flow regime $\gamma >$ 200 %). The split of the fluid in opposite temperature variation consists of cold and hot regions aligned along the strain and a thermal compensation between the generated bands. The hot band is located close to the moving plate while the cold band is above the hot bands. The maximum temperature variation amplitude was reached at large strain (2500%) reaching $|\Delta T| = 0.04 \pm 0.01$°C (Figure 2).

Since the lowest strain amplitudes correspond to less perturbing conditions, the elastic response is the fundamental response of blood plasma. No thermal response is measurable: the temperature is rather homogeneous. This is consistent with a conservative (elastic) mechanism. The shear elasticity becomes progressively replaced by a viscous response as shear strains rise. The fluid becomes thermally heterogeneous, whereas the viscoelastic measurement indicates a viscous regime only (G" >> G'). A viscous regime is incompatible with a thermal inhomogeneity. Consequently, the apparent viscous behavior is the non-linear product of the mechanical action on the shear elasticity converted in local thermodynamic states. Similar strain dependence schemes were reported in liquids and viscoelastic fluids (polymer melts, glass formers) by different authors[5-9,11-14,19-20]. Similar thermo-elastic effects were identified in simple fluids (glycerol, polypropylene glycol) confirming that the elasticity is likely a generic property[15,27].

Strain amplitude sweep tests carried out at several sample thicknesses show that the elastic response is accessible up to 1.100 mm using optimized wetting conditions, but decreases with increasing gap width (Figure 3). The shear elastic property is therefore lost at high strain amplitudes and large thicknesses; i.e. away from equilibrium conditions. Figure 3 thus summarizes the evolution of the shear elasticity of blood plasma as a function of its thickness. The insert of Figure 3 shows the $G'$-values at near-linear response (elastic plateau at low strain amplitude). Like explained above, the elastic plateau developed at very low strains ($\gamma < 5$ %) and decreased until both moduli dropped abruptly at about 300 %, at which strain the sample yielded (compare also with Figure 2). After yielding, the shear elastic moduli converged and became indistinguishable from values measured at other gap thicknesses.

Discussion and Conclusion

We have revealed that blood plasma exhibited a finite shear elastic response within a frequency range that covers the human heart rate if the interfacial contact is optimized. Its response to an oscillatory shear stress excitation in nearly static conditions (down to 0.04 rad/s upon applying low strain amplitude perturbation) is fundamentally elastic-like. Thus, for blood plasma to flow, the strain impulse has to overpass an elastic threshold. Significant shear strains and pulsations are needed to overcome the fundamental elastic property of blood plasma. Our results support recent findings, showing that blood plasma is viscoelastic when exposed to extensional deformation[3,4]. Additionally we have demonstrated the dimensional properties of shear elasticity by highlighting its reinforcement by the elastic threshold when the probed volume was reduced. Reciprocally, the elastic behavior is progressively replaced at large scale and/or at large strain amplitudes by a flow regime.

The present study evidences that a modification of the boundary surface has deep consequences for flow. Since the scale and the adhesion of blood plasma to the wall determine its strain-dependent behavior, the physico-chemistry of the interfacial boundary highly affects the volume flow through channels. An immobile plasma layer (not only at vessel walls) diminishes the available space for blood transport. It might even influence the material transport towards the wall and through it by selective binding. To optimize biomedical devices (e.g., extracorporeal membrane oxygenation, hemodialysis systems) and protocols (e.g., such as model blood flow through such devices), controlled interfacial conditions and true dimensional scales that reflect the planned application are key parameters. Since the threshold for flow is low, the boundary effect will be less significant at high flows.

In blood vessels, the role of plasma for the boundary surface is essential. The immersion of the endothelial glycocalix in plasma generates a stabile hydrogel layer (endothelial glycocalix layer, EGL) of several hundreds' nanometers in width that sieves molecules, hinders excessive fluid flux into the interstitium[28-31], and limits the microvascular perfusion heterogeneity[31]. Therefore, flow does not start immediately at the endothelial cell bilayer but at a distance. Measurements of the flow resistance estimate this distance in the micrometer range from the endothelial cell[33]. The immobile part of the cell-free layer in vessels is therefore larger than the EGL width. The role of plasma is fundamental in this context because it contains the same main components as the glycocalix layer and can couple perfectly to it. Reitsma and co-authors described the surface layer as highly dynamic due to the synergetic interaction of soluble plasma components with the membrane bound components of the glycocalix[34]. The interplay between the two layers is disrupted when one part is degraded, which is the case at enzymatic treatment[35].

The identification of the elastic behavior of blood plasma at low strain (Fig.1-3) reveals thus the dynamic range for which the EGL might be in full interaction with the plasma. The elasticity was highlighted by optimizing the conditions of the stress transfer that appear as playing a key role in the dynamic characterization of fluids[5-9,14-15]. "Static" shear elasticity reveals collective properties that involve long-range molecular like hydrogen bond interconnectivity. The thermal heterogeneity at large strain is also coherent with a thermo-mechanical coupling only possible through the deformation of a pre-existent shear elasticity as already observed in simple fluids and confirming that the elasticity is likely a generic fluidic property[5-9],[15]. Blood plasma definitively cannot be treated as a viscous fluid; this assumption strongly underestimates its properties and physiological relevance.

Limitations: We propose an optimized system but since the walls are rigid, it can neither take into account the elasticity of the vessel wall (pulse pressure), nor the deformation of the glycocalix layer. Also, while the plasma is stabilized, long time scale experiments (hours) are however not possible, a crust being created on the fluid surface, especially at low gap widths. The thermal study is also limited by a poor spatial resolution (20µm) and by a limited dynamics. A high frequency recording of thermal images is hardly accessible but not necessary in the frame of physiological conditions.

| Parameter | Tested range |
|---|---|
| Frequency $\omega$ (rad/s) | 0.05 rad/s to 100 rad/s |
| Strain amplitude $\gamma$ (%) | 1 to 2000% |
| Gap thickness $e$ (mm) | 300 µm up to 1150 µm |
| Contact angle ($\theta$) | Total wetting $\theta \approx 0$ (macroscopically) |

**Table 1**

Table 1 indicates parameters varied for the study and the ranges of their use. These settings were chosen on basis of anatomical and physiological circumstances. The gap thickness of 300 µm reflects the diameter of a large arteriole, whereas 1100 µm reflects the diameter of a medium artery. In these resistance vessels the immobile part of the cell-free plasma layer is most relevant to blood flow. The physiologic heart rate covers 6 (at rest) and 19 rad/s (during exercise). This frequency range was included into the frequency sweep tests (Figure 1). Shear strain amplitudes differ across the vessel cross-section, being zero in the center of the flow, and maximal above the endothelial surface layer. Mean arterial wall shear rates are in the range of 300-500 $s^{-1}$, capillary wall shear rates are higher than 1000 $s^{-1}$ [36]. Thus, the strain amplitude of 1 % reflects conditions in the central part of the flow profile whereas the high strain amplitude (2000 %) reflect conditions in its parabolic part. It should also be taken into account that within the glycocalyx layer, to which blood plasma is attached, there may be deformation amplitudes of 1.5% [37].

**Author's contributions:** all authors contribute equally to this work.

**Data availability:** The data that support the findings of this study are available from the corresponding author upon reasonable request.

Figures and captions:

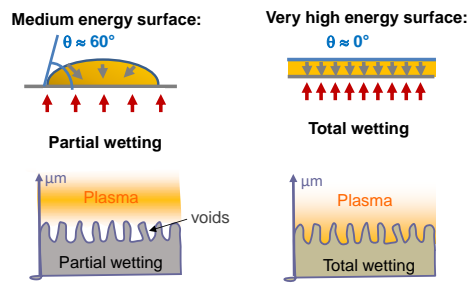

**Scheme 1.** Differences at macroscopic scale (upper schemes representing the different drop behaviors) and micron-scale (bottom schemes) between partial wetting (scheme showing an about 60° contact angle) (a) and total wetting reaching zero contact angle (b) mechanisms. High-energy surfaces are required to provide surface boundary conditions in line with physiological interacting fluids.

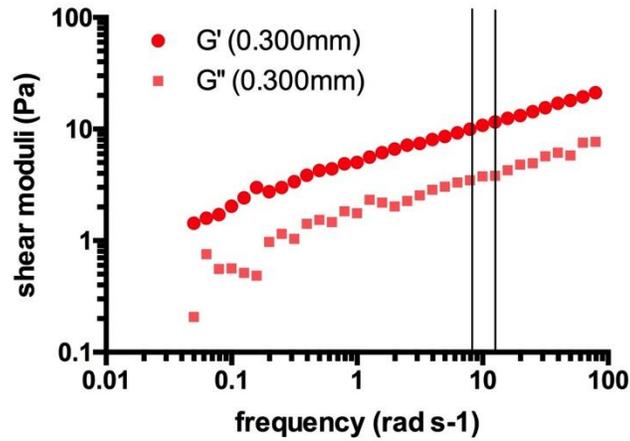

**Figure 1.** Measurements in well-controlled wetting (alumina surfaces): dynamic moduli (*G'* (●) and *G''* (■)) of a 0.300mm thick blood plasma versus frequency in low strain conditions (total wetting conditions - 1.5 % strain amplitude, room temperature measurements). The vertical lines point out the heart rate domain (1 - 1.5 Hz). The error bar is about 5% of the measured values (logarithmic scale).

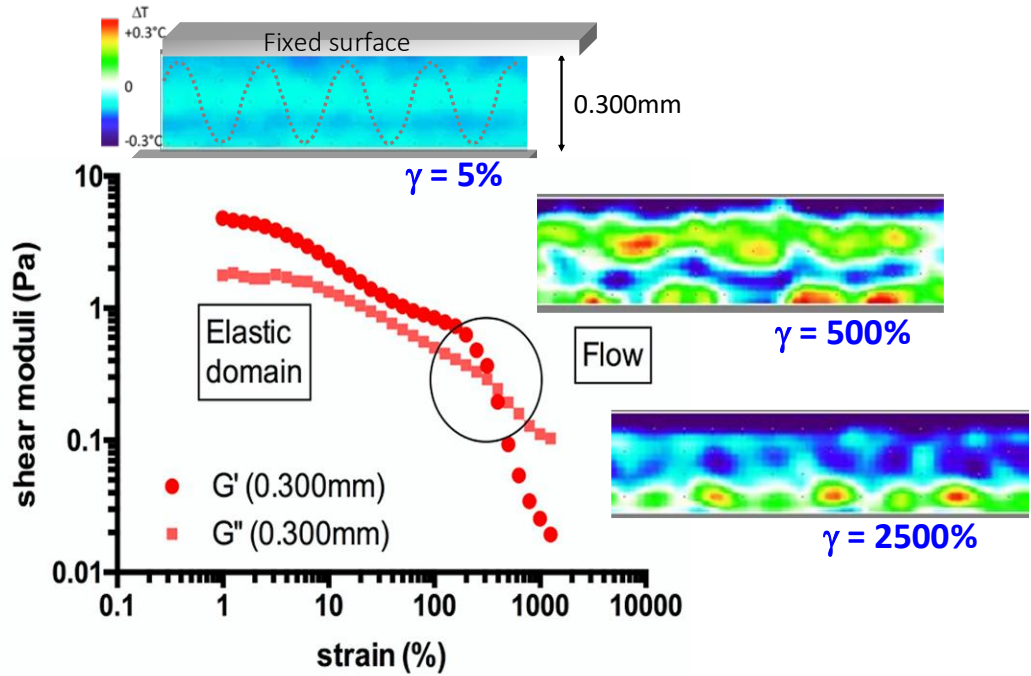

**Figure 2.** Human blood plasma exhibits, in the heart rate domain (here $\omega = 10$ rad/s), an elastic-like response at low strain, the elastic modulus $G'$ (●) being larger than $G''$ (■) for $\gamma < 200\%$. At larger deformation, a strain-induced transition from elastic to apparent viscous regime is observed and that is simultaneous with the emergence of contrasted thermal zones of sizeable amplitude ± 0.2°C (*in situ* microthermal cartography in the gap view). Measurements carried out at 0.300mm gap thickness using total wetting conditions. Due to instrumental limitation, higher strain amplitudes were not probed ($\gamma_0$ evolves from 1 % to 1200 %). The error bar is about 5% of the measured values (logarithmic scale).

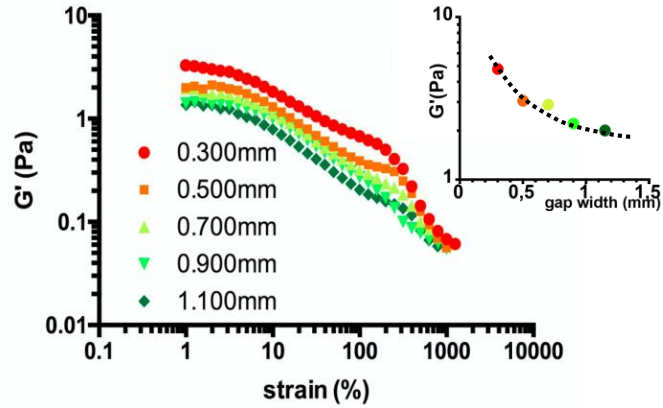

**Figure 3.** Influence of the scale (gap thickness) on the blood plasma shear elasticity measured at different thicknesses from 0.300 mm up to 1.100 mm versus strain amplitude (room temperature measurements, total wetting (alumina surfaces)): ●: 0.300 mm, ■: 0.500 mm, ▲: 0.700 mm, ▼: 0.900 mm, ♦: 1.150 mm. The insert summarizes the evolution of the low strain amplitude shear elasticity (values near to the elastic plateau, logarithmic scale) versus gap thickness; measurements in high wetting conditions (alumina surfaces). The error bar is about 5% of the measured values (logarithmic scale).